\documentclass[useAMS,usegraphicx,usenatbib,letterpaper]{emulateapj}

\usepackage{ulem}
\usepackage{color}
\usepackage{graphics,graphicx}
\usepackage{times} 
\usepackage{amssymb}
\usepackage{amsmath}
\usepackage{natbib}
\usepackage{url}
\usepackage{multirow}
\usepackage{ctable}
\usepackage{threeparttable}
\newif\ifAMStwofonts
\AMStwofontstrue

\def\xmm{{\it XMM-Newton}}

\def\suzaku{{\it Suzaku}}

\def\epicmos1{{\it EPIC}{\rm-MOS1~\/}}
\def\epicmos2{{\it EPIC}{\rm-MOS2 ~\/}}
\def\epicmos{{\it EPIC}{\rm-MOS}}
\def\xis{{\rm XIS}}
\def\pin{{\rm PIN}}
\def\hxd{{\rm HXD}}
\def\nustar{{\it NuSTAR}}

\def\deg{$^{\circ}$}

\def\H0{{\rm ~km~s^{-1}~Mpc^{-1}}}

\def\kev{\hbox{\rm keV}}

\def\atpcm{{\rm atom~cm$^{-2}$}}

\def\ergpcmsqps{\hbox{$\rm\thinspace erg~cm^{-2}~s^{-1}$}}

\def\ergcmps{\hbox{\rm erg~cm~s$^{-1}$}}

\def\msun{\hbox{$\rm\thinspace M_{\odot}$}}

\def\chisq{{$\chi^{2}$}}

\def\xspec{\hbox{\footnotesize XSPEC}}

\def\heasoft{\hbox{\rm{\footnotesize HEASOFT}}}
\def\xselect{\hbox{\rm{\footnotesize XSELECT}}}

\def\specgroup{\hbox{\rm{\footnotesize SPECGROUP}}}

\def\addascaspec{\hbox{\rm{\footnotesize ADDASCASPEC}}}
\def\sas{\hbox{\rm{\footnotesize SAS}}}

\def\xisresp{\hbox{\rm{\footnotesize XISRESP}}}

\def\hxdpinxbpi{\hbox{\rm{\footnotesize HXDPINXBPI}}}

\def\grid25{\hbox{\rm{\footnotesize GRID25}}}

\def\tbnew{\rm{\footnotesize TBNEW}}

\def\reflionx{\rm{\footnotesize REFLIONX}}

\def\gsmooth{\rm{\footnotesize GSMOOTH}}

\def\fexxvi{\hbox{\rm Fe\,{\small XXVI}}}

\def\eg{{\it e.g.~\/}}

\def\ie{{\it i.e.~\/}}

\def\la{\mathrel{\hbox{\rlap{\hbox{\lower4pt\hbox{$\sim$}}}{\raise2pt\hbox{$<$}}}}}
\def\ga{\mathrel{\hbox{\rlap{\hbox{\lower4pt\hbox{$\sim$}}}{\raise2pt\hbox{$>$}}}}}

\def\d25{D$_{25}$}

\def\.25{0.25 keV\thinspace}

\def\suzhxdlink{http://www.astro.isas.ac.jp/suzaku/analysis/hxd/}

\def\ngc{NGC\,6814}

\shorttitle{Hard X-ray Lags in AGN: NGC\,6814}
\shortauthors{D.~J. Walton et al.}

\begin{document}

\title{Hard X-ray Lags in Active Galactic Nuclei: Testing the Distant Reverberation
Hypothesis with NGC 6814}

\author{D. J. Walton\altaffilmark{1},
A. Zoghbi\altaffilmark{2,3},
E. M. Cackett\altaffilmark{4},
P. Uttley\altaffilmark{5},
F. A. Harrison\altaffilmark{1},
A. C. Fabian\altaffilmark{6},
E. Kara\altaffilmark{6},
J. M. Miller\altaffilmark{7},
R. C. Reis\altaffilmark{7},
C. S. Reynolds\altaffilmark{2,3}}
\affil{
$^{1}$Cahill Centre for Astronomy and Astrophysics, California Institute of
Technology, Pasadena, CA 91125, USA \\
$^{2}$Department of Astronomy, University of Maryland, College Park, MD 20742,
USA \\
$^{3}$Joint Space-Science Institute (JSI), University of Maryland, College Park, MD 20742, USA \\
$^{4}$Department of Physics and Astronomy, Wayne State University, Detroit, MI
48201, USA \\
$^{5}$Astronomical Institute ‘Anton Pannekoek’, University of Amsterdam, Postbus
94249, NL-1090 GE Amsterdam, the Netherlands \\
$^{6}$Institute of Astronomy, University of Cambridge, Madingley Road, Cambridge,
CB3 0HA, UK \\
$^{7}$Department of Astronomy, University of Michigan, 500 Church Street, Ann
Arbor, MI 48109, USA
}

\begin{abstract}
We present an X-ray spectral and temporal analysis of the variable active galaxy
\ngc, observed with \suzaku\ during November 2011. Remarkably, the X-ray
spectrum shows no evidence for the soft excess commonly observed amongst
other active galaxies, despite its relatively low level of obscuration, and is
dominated across the whole \suzaku\ bandpass by the intrinsic powerlaw-like
continuum. Despite this, we clearly detect the presence of a low frequency hard
lag of $\sim$1600s between the 0.5--2.0 and 2.0--5.0\,\kev\ energy bands at
greater than 6$\sigma$ significance, similar to those reported in the literature
for a variety of other AGN. At these energies, any additional emission from \eg
a very weak, undetected soft excess, or from distant reflection must contribute
less than 3\% of the observed countrates (at 90\% confidence). Given the lack of
any significant continuum emission component other than the powerlaw, we can
rule out models that invoke distant reprocessing for the observed lag behavior,
which must instead be associated with this continuum emission. These results
are fully consistent with a propagating fluctuation origin for the low
frequency hard lags, and with the interpretation of the high frequency soft lags
-- a common feature seen in the highest quality AGN data \textit{with} strong
soft excesses -- as reverberation from the inner accretion disk.
\end{abstract}

\section{Introduction}

Physical interpretation of the complex X-ray spectral and temporal behavior of
active galaxies has been subject to continuous debate. One of the most popular
interpretations invokes relativistic disc reflection (\citealt{FabZog09,
Walton10hex, Brenneman11, Reis3783, Nardini12, Gallo13}). In this scenario,
the powerlaw-like coronal emission irradiates the accretion disk, resulting in
an additional `reflected' emission component consisting of both continuum
emission and discrete atomic features, the most prominent of which is typically
the iron K$\alpha$ emission line (\citealt{George91}). The majority of the
emission arises from the inner regions of the accretion flow, and the relativistic 
effects inherent to such regions of strong gravity can broaden and blend these
atomic features to the extent that at soft energies the reflected emission appears
as an additional smooth continuum component (the `soft excess';
\citealt{Crummy06, Nardini11, Walton13spin}), and result in the characteristic
broad, skewed relativistic iron emission line profile first observed in
MCG--6-30-15 (\citealt{Tanaka95}). Comparison with the spectra observed from 
Galactic black hole binaries (BHBs) appears to support the presence of relativistic
disc reflection in active galaxies (\citealt{Walton12xrbAGN}), and indeed this has
now been strongly confirmed for the variable AGN NGC1365
(\citealt{Risaliti13nat}), through coordinated observations with \xmm\ and the
recently launched \nustar\ observatory (\citealt{NUSTAR}).

Relativistic disc reflection also offers a natural explanation for some of the
temporal behavior typically observed, in particular the frequency dependent time
lags (the `lag frequency-spectrum') between different energy bands. Much of the
work in this field has naturally focused on reverberation of the soft excess, owing
to the photon statistics required for detailed lag studies, which on short
timescales is observed to lag behind the emission at higher energies dominated 
by the intrinsic powerlaw continuum (`soft' lags; \citealt{FabZog09,
Zoghbi11, deMarco13, Fabian13iras, Cackett13}), a natural expectation of disc
reflection. More recently, similar high-frequency reverberation of the broad iron
K line has also been detected in sufficiently bright AGN (\citealt{Zoghbi12,
Zoghbi13a, Kara13iras, Kara13feK}). However, in contrast to this apparent
short-timescale reverberation, for variations on long timescales, higher energy
emission is observed to lag behind that at lower energies (`hard' lags;
\citealt{McHardy04, Papadakis01, Vaughan03}). In the reflection scenario, these
low-frequency hard lags are believed to be intrinsic to the powerlaw-like
continuum rather than related to the reflection process. Similar hard lags are
seen in Galactic BHBs, and are currently generally believed to arise due to
inwardly propagating mass accretion rate fluctuations (\citealt{Kotov01,
Arevalo06}), although an origin related to Compton scattering, widely expected
to produce the powerlaw continuum, has also been discussed in the literature
(\citealt{Kazanas97}).

\begin{figure*}
\hspace*{-0.4cm}
\epsscale{0.54}
\plotone{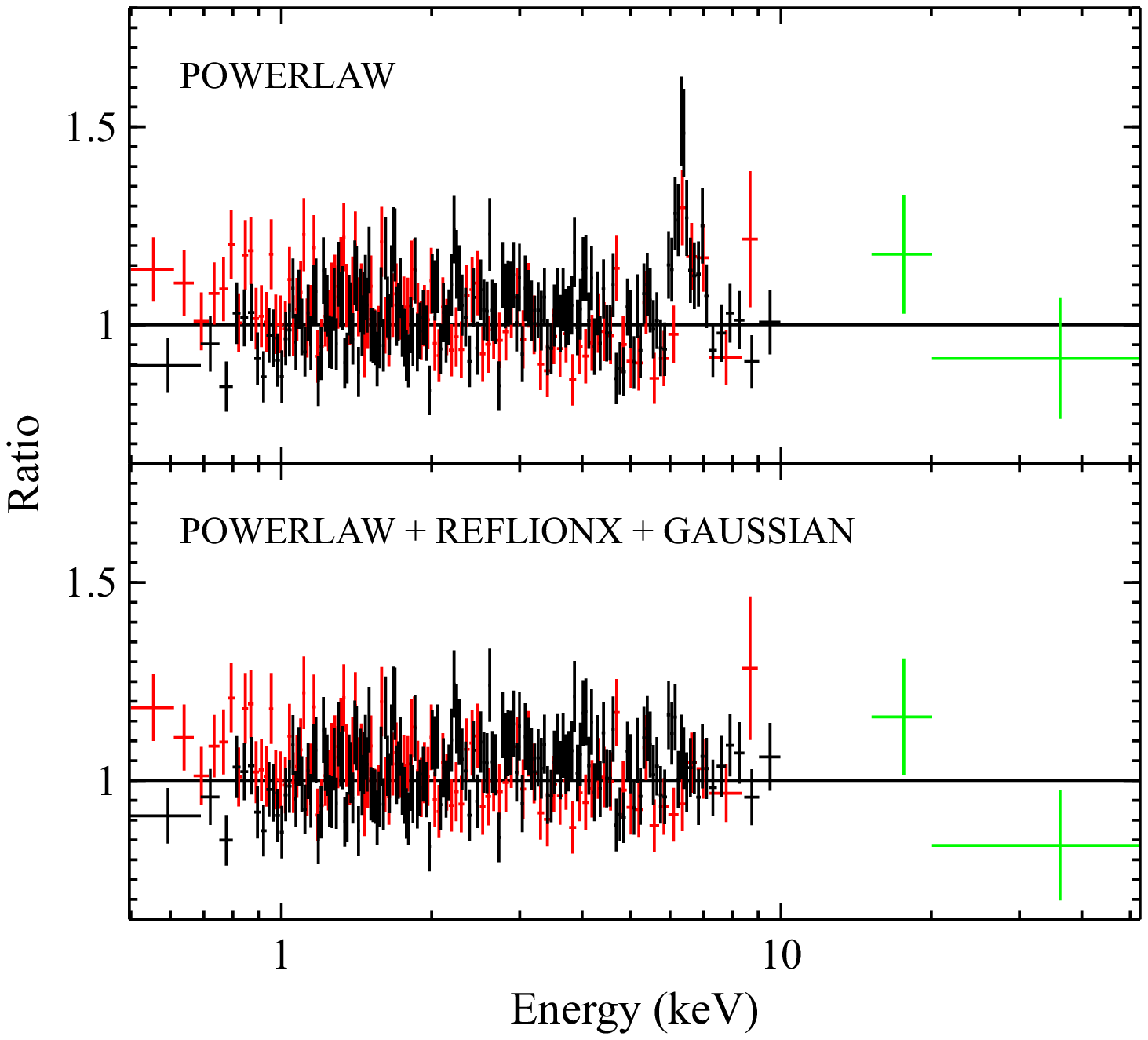}
\hspace*{1.0cm}
\plotone{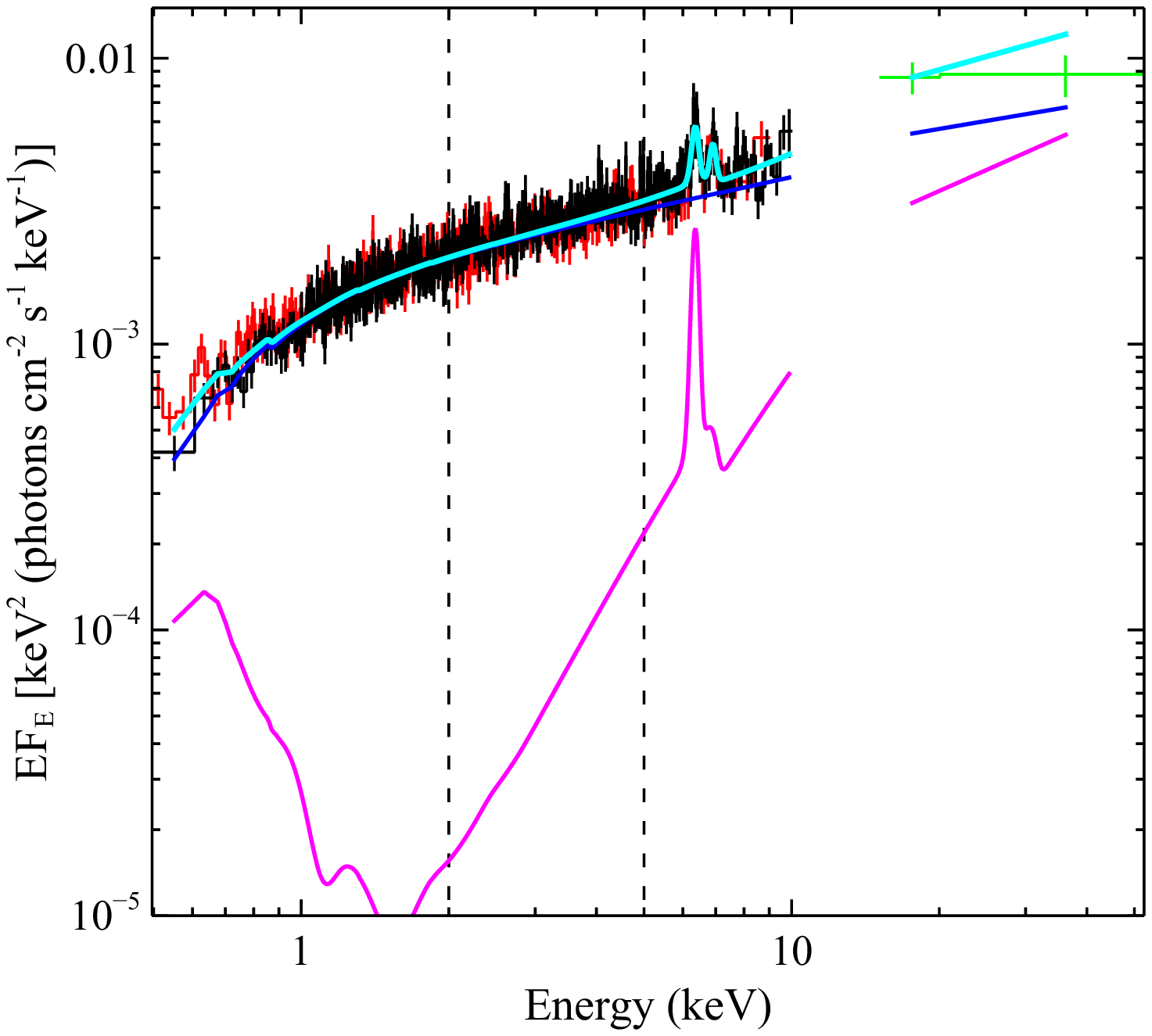}
\caption{
\textit{Left:} data/model ratios for a simple absorbed powerlaw, and a powerlaw
with a distant reflector and a Gaussian emission line. \textit{Right:} the relative
contributions of the powerlaw (blue) and distant reflection (magenta), the latter
indicating the upper limit obtained for the reflection in the marked 2--5\,\kev\
bandpass. XIS FI, BI and PIN data are shown in black, red and green, and the total
model in cyan; spectra have been rebinned and the additional Gaussian omitted
from display for visual clarity. The broadband spectrum is dominated by the
intrinsic powerlaw continuum.}
\label{fig_spec}
\end{figure*}

However, it has recently been argued that both the hard lags and the apparent
reverberation of the soft excess instead arise simultaneously through
reverberation from more distant optically thick material, rather than via separate
physical processes (\citealt{LMiller10b, Legg12}). In this scenario, the
low-frequency hard lags are interpreted as the true signature of reverberation,
while the high-frequency soft lags are proposed to arise as a secondary
consequence of this same reverberation process. This requires the reverberation
transfer function -- which determines how the hard band is observed to respond 
to variations in the soft band -- to have a sharp functional form, resulting in
oscillatory behavior in the lag frequency-spectrum at high frequencies (``phase
wrapping''), which if strong enough (or the data  sensitive enough) can
additionally result in the apparent observation of soft lags at high frequencies. In
this scenario strong non-zero structure should only be observed in lag spectra
when comparing energy bands in which at least one contains a substantial
contribution due to reprocessing by distant absorbing material. Such additional
absorption/emission will naturally serve to distort the observed energy spectrum
away from the intrinsic powerlaw-like AGN continuum. This interpretation for the
lag behavior is designed to complement the absorption dominated solutions for
the observed X-ray spectral complexities (\citealt{LMiller09, Turner09}), although
at the time of writing interpretations that self-consistently explain both the
temporal and spectral behavior have not yet been presented.

Here, we present a new analysis of a recent \suzaku\ observation of the variable
AGN \ngc\ (\citealt{Mukai03}), in which we demonstrate that the hard lags are
most likely intrinsic to the powerlaw-like continuum, rather than arising through 
reverberation from distant circumnuclear material.

\section{Data Reduction}
\label{sec_red}

\suzaku\ (\citealt{SUZAKU}) observed \ngc\ in November 2011 for $\sim$85\,ks
duration. Here, we are primarily interested in the data obtained with the XIS
detectors (\citealt{SUZAKU_XIS}). Following the \suzaku\ data reduction
guide\footnote{http://heasarc.gsfc.nasa.gov/docs/suzaku/analysis/}, we
reprocessed the unfiltered event files for each of the \xis\ CCDs (XIS0, 1, 3) and
editing modes (3x3, 5x5) operational using \heasoft\ (v6.13), re-running the
pipeline with the latest calibration database release (Nov 2012) and
the standard screening criteria. Source products were taken from circular
regions $\sim$200'' in radius, and background estimated from regions free of any
contaminating sources in the surrounding areas on the CCD. Spectra and
lightcurves were extracted from the cleaned event files with \xselect, and
responses were generated for each individual spectrum using the \xisresp\
script (with medium resolution). After confirming their consistency, the spectra
and responses for the front-illuminated (FI) detectors (XIS0 and 3) were combined
using \addascaspec. Finally, we grouped the spectra to have a minimum
signal-to-noise (S/N) of 5 per energy bin with \specgroup\ (part of the \xmm\
\sas), to allow the use of \chisq\ minimization during spectral fitting. 

For the collimating \hxd\ \pin\ detector (\citealt{SUZAKU_HXD}) we again
reprocessed the unfiltered event files. Estimation of the background requires
individual consideration of the non X-ray instrumental background (NXB; the
dominant component) and the cosmic X-ray background (CXB). Although we
initially considered the recommended `tuned' NXB model provided by the
\suzaku\ team, doing so resulted in a severe under-prediction of the
cross-normalization between the XIS and PIN data for any reasonable model in
this case, giving $C_{\rm{PIN/XIS}}\sim0.5$ rather than the expected $\sim$1.17
(the opposite effect a hard excess would give). Instead we used the earth-occulted
data (earth elevation angle $<-5$\deg) as the NXB contribution, which gave
much more sensible results with the standard cross-calibration. The PIN response
matrix was downloaded for the relevant observing epoch\footnote{\suzhxdlink}.
Spectral products were generated using the \hxdpinxbpi\ script, which simulates
the expected contribution from the CXB, and the source spectrum was re-binned
to a minimum S/N of 3 per bin, lower than required in the XIS spectra owing to the
much lower countrate, but still sufficient to allow the use of \chisq\ minimization. 

During this work, we consider the XIS data over the 0.5--10.0\,\kev\ energy
range, and the PIN data over the 15--50\,\kev\ energy range. The data from all
the detectors are modeled simultaneously, with all parameters tied between the
spectra, and we attempt to account for any cross-calibration uncertainties
between the FI and BI XIS detectors by allowing a multiplicative constant to vary
between them. This value is always found to be within 1 per\,cent of unity. Unless
stated otherwise, quoted uncertainties on spectral parameters are the 90\%
confidence limits throughout, and spectral analysis is performed with \xspec\
v12.8.0 (\citealt{XSPEC}).

\begin{figure}
\epsscale{1.1}
\hspace{-0.25cm}
\plotone{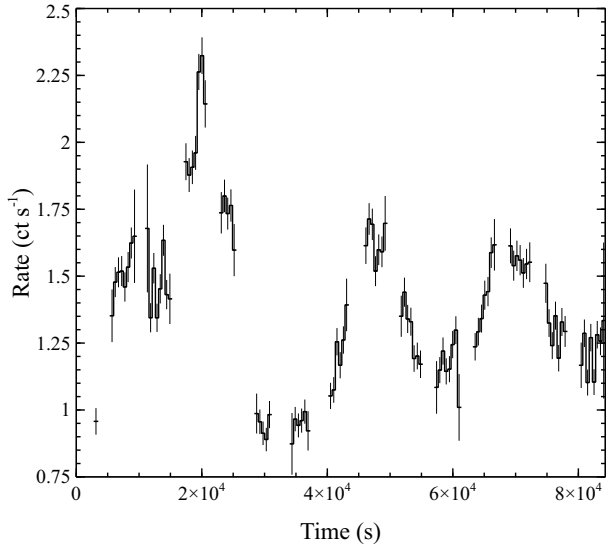}
\caption{The combined 0.5--10\,\kev\ XIS lightcurve for \ngc, rebinned to 512s
for clarity, showing the variability displayed during this \suzaku\ observation.}
\label{fig_lc}
\end{figure}

\section{Spectral Analysis}
\label{sec_spec}

We begin by modeling the 0.5--50.0\,\kev\ spectrum of \ngc\ with a simple
absorbed powerlaw, allowing for both Galactic absorption
($N_{\rm{H;Gal}}=8.98\times10^{20}$\,\atpcm; \citealt{NH}) and intrinsic neutral
absorption, utilising the \tbnew\ absorption
code\footnote{http://pulsar.sternwarte.uni-erlangen.de/wilms/research/tbabs}
along with the solar abundances quoted in \cite{tbabs}. Aside from some residual 
features at the iron K energies, this simple model provides an extremely good fit
to the \suzaku\ data (see Fig. \ref{fig_spec}); the statistics are naturally dominated
by the XIS data, but with the standard cross-calibration ($C_{\rm{PIN/XIS}}=1.17$)
the PIN spectrum is fully consistent with an extrapolation of the lower energy data. 
The photon index is fairly hard, $\Gamma=1.53\pm0.02$, however \ngc\ is radio
quiet (\citealt{Xu99}), so the X-ray spectrum should not be jet-dominated, and the
amount of intrinsic neutral absorption is very small,
$N_{\rm{H;int}}=3.4\pm1.5\times 10^{20}$\,\atpcm.

The Fe K residuals show clear evidence for emission from neutral iron. Including
a Gaussian emission line improves the fit by $\Delta\chi^{2}=76$ for 3 additional
free parameters. There may also be emission associated with hydrogen-like iron
at $\sim$7\,\kev. Including a second line provides only a moderate improvement,
$\Delta\chi^{2}=18$ (3 additional free parameters), but doing so improves the 
constraints on the neutral line parameters, which no longer also attempt to
account for the weaker excess emission at higher energies. Therefore, although
the significance of this second line is more marginal, we include it in all our
subsequent fits. The line energies obtained are $E_{\rm{FeI}}=6.40\pm0.03$ and
$E_{\rm{FeXXVI}}=6.94\pm0.07$\,\kev. We find the widths of the two lines to be
consistent with one another, so for simplicity we link these parameters, obtaining
$\sigma=0.12\pm0.04$\,\kev, and their equivalent widths are
$EW_{\rm{FeI}}=170^{+30}_{-40}$ and $EW_{\rm{FeXXVI}}=90^{+30}_{-40}$\,eV.
The resulting fit is excellent, with $\chi^{2}_{\nu}=1398/1371$.

\begin{figure}
\epsscale{1.1}
\hspace{-0.25cm}
\plotone{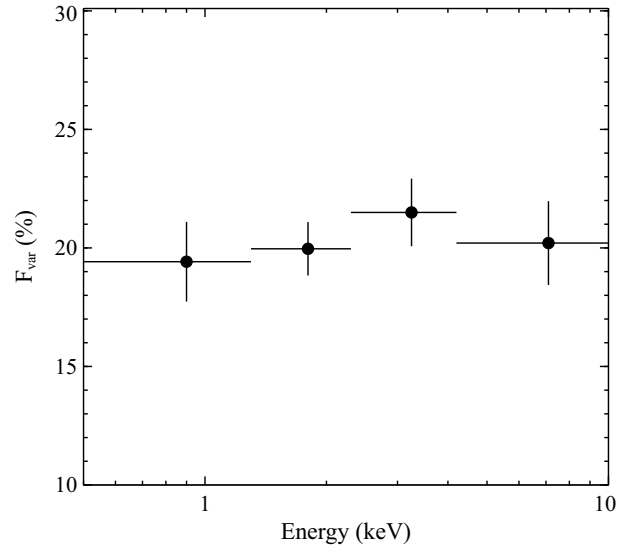}
\caption{The fractional excess variability ($F_{\rm{var}}$; covering frequencies
$\sim$1.2$-200\times10^{-5}$\,Hz) as a function of energy. No significant
energy dependence is observed, as expected if the variability is dominated by a
single emission component with a roughly constant spectral shape.}
\label{fig_fvar}
\end{figure}

Critically, the \suzaku\ spectrum of \ngc\ does \textit{not} display the soft
excess typically observed in other unobscured AGN; in terms of the continuum
the observed spectrum is dominated by the intrinsic powerlaw-like emission.
Addition of a cool blackbody component, frequently used to phenomenologically
model the soft excess, does not improve the fit ($\Delta\chi^{2}=3,2$
additional free parameters). For a temperature of 0.15\,\kev, as is typically
obtained (\citealt{Gierlinski04, Miniutti09}), we find the upper limit to the
0.5--2.0\,\kev\ blackbody flux to be less than 3\% of the powerlaw flux in this
bandpass (at 90\% confidence), even allowing for a variable intrinsic absorption
column. When present in other AGN, the soft excess usually contributes
$\sim$50\% of the 0.5--2.0\,\kev\ powerlaw flux when modeled with a
blackbody component (\citealt{Miniutti09}). In fact, for many of the particular
AGN on which the debate over the origin of the spectro-temporal behavior has
focused, \eg 1H\,0707-495, the soft excess is substantially stronger than
average.

The narrow iron line may indicate the presence of reflection by distant material.
We therefore replace the neutral Gaussian with a full reflected spectrum (\reflionx;
\citealt{reflion}). Based on the line widths obtained previously, we allow some mild,
Gaussian broadening with \gsmooth, with the width linked to the \fexxvi\ line.
Again, an excellent fit is obtained, with $\chi^{2}_{\nu}=1397/1370$. The iron
abundance is constrained to be $A_{\rm{Fe}}\rm{/solar}>1.0$, the ionisation of
the reflector is low, $\xi<60$\,\ergcmps, and the line broadening is consistent
with that obtained previously: $\sigma=0.11\pm0.04$\,\kev. This reflection
component is weak ($R<0.6$, calculated following \citealt{Walton13spin}) and
contributes less than 2--3\% of the observed countrates in each of the 0.5--2.0
and 2.0--5.0\,\kev\ bands, given the XIS responses. The observed 2--10\,\kev\
flux from \ngc\ is $8.2\pm0.1\times10^{-12}$\,\ergpcmsqps, which for a
distance of 18.5\,Mpc, mass of $2\times10^{7}$\,\msun\ (\citealt{Bentz09}) and
the appropriate bolometric correction (\citealt{Vasudevan09}) implies a rather low
Eddingtion ratio, $L_{\rm{Bol}}/L_{\rm{Edd}}\sim10^{-3}$. In this regime, the inner
disk may be truncated. Alternatively, the corona could be mildly outflowing
(\citealt{Beloborodov99}). Either would offer a natural explanation for the weak
reflection and the lack of a soft excess, and the latter could also explain the hard
spectrum.

\begin{figure}
\epsscale{1.10}
\hspace*{-0.5cm}
\plotone{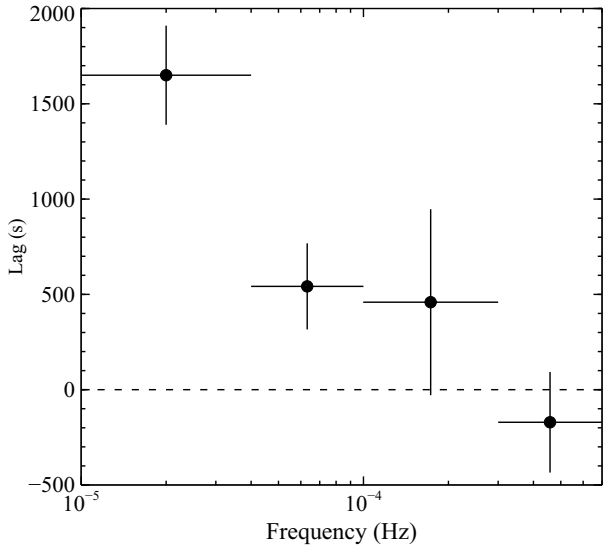}
\caption{The lag frequency-spectrum obtained between the 0.5--2.0 and
2.0--5.0\ \kev\ energy ranges. At low frequencies, a
hard lag of $\sim$1600\,s is clearly detected, at greater than 6$\sigma$
significance.}
\label{fig_lagFspec}
\end{figure}

\section{Variability and Lags}
\label{sec_lag}

Given the unusual lack of the spectral features that have been so hotly debated
in the literature, this observation of \ngc\ provides the ideal testing ground to
distinguish between some of the varying explanations for the hard lags seen in
other AGN. In Fig. \ref{fig_lc} we show the combined 0.5--10.0\,\kev\ XIS
lightcurve, visibly demonstrating the variability displayed during this epoch. 
Despite this clear flux variability, the spectrum retains its simple powerlaw-like
shape throughout. The fractional excess variability (\citealt{Edelson02,
Vaughan03}) for full lightcurve (covering frequencies
$\sim$1.2$-200\times10^{-5}$\,Hz) does not display any significant energy 
dependence (see Fig. \ref{fig_fvar}), as expected if the flux variability is
dominated by a single component that does not strongly vary its spectral shape.

We therefore investigate the presence/behavior of any time lags displayed by
\ngc\ during this period, in order to provide a comparison with sources that
display more complex X-ray spectra. To calculate the lag frequency-spectrum,
we utilize the method introduced in \cite{LMiller10b}, and detailed in
\cite{Zoghbi13b}, which is designed specifically for use with windowed time
series, a result of the low-earth orbit of \suzaku. In brief, the method relies on
maximizing the likelihood function. The model of the likelihood function is
constructed from the cross-correlation function, whose input parameters are
the lag values in pre-defined frequency bins (four in this case, giving the best
balance between S/N and frequency resolution). The lag frequency-spectrum is
inverse-Fourier transformed to give the cross-correlation function, which in
turn is used to construct the time-domain covariance matrix of the light curve
values, and hence the probability of obtaining the data given the model (\ie the
likelihood). The final result is then obtained by maximizing this likelihood. The
energy ranges used here are 0.5--2.0 and 2.0--5.0\,\kev\ for the soft and
hard bands, comparable to similar studies in the literature.

\begin{figure}
\epsscale{1.10}
\hspace*{-0.5cm}
\plotone{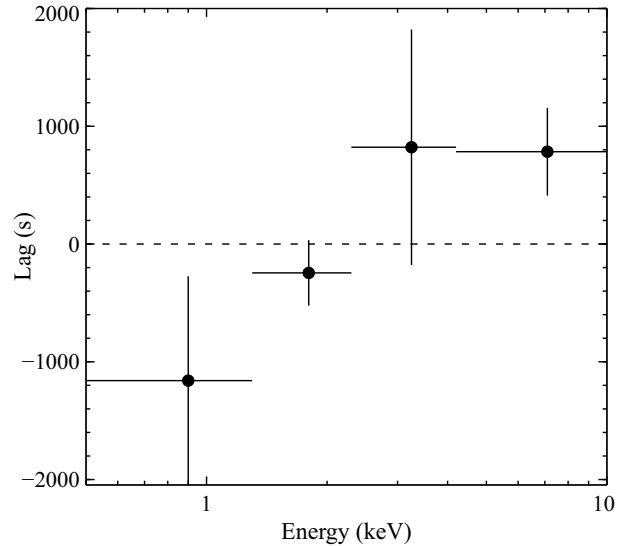}
\caption{The lag energy-spectrum obtained during this epoch, computed over
the frequency range $1-20\times10^{-5}$\,Hz. The lag relative to the
softest energies increases with increasing energy.}
\label{fig_lagEspec}
\end{figure}

The lag frequency-spectrum obtained is shown in Fig. \ref{fig_lagFspec}. At the
lowest frequencies probed, the 2.0--5.0\,\kev\ emission lags behind the softer
0.5--2.0\,kev\ emission by $1650\pm260$s. The detection of the hard lag is
highly significant: the lowest frequency bin alone is inconsistent with zero lag at
greater than the 6$\sigma$ level based on the obtained uncertainties, and the
total deviation from zero lag towards lower frequencies is more than 8$\sigma$.
A model of zero lag provides a very poor fit, with $\chi^{2}_{\nu}=47/4$ (null
hypothesis probability of $p\sim5.3\times10^{-11}$; lag errors are
$\sim$Gaussian). We also show the lag energy-spectrum in Fig. \ref{fig_lagEspec},
computed over the frequency range $1-20\times10^{-5}$\,Hz and using the full
0.5--10.0\,\kev\ energy range as the reference band, which results in an offset
of the lags obtained, but does not change the relative behavior. The lag relative
to the softest band increases with increasing energy. All of this behavior is
consistent with that reported for a variety of other AGN (\citealt{McHardy04,
Papadakis01, Vaughan03, Kara13feK}). Although the spectrum does not vary
strongly, some weak pivoting must occur to produce the lags (\citealt{Kording04}).

Finally, given our spectral results, we test whether distant reprocessing could
produce the observed lags. Following \citealt{LMiller10b}, we model the observed
lag frequency-spectrum as arising from a simple tophat reverberation transfer
function for the reflection, plus a $t=0$ delta-function for the continuum, both
of which naturally contribute to the flux in both bands considered. The fractional
contribution to the total hard band countrate from the reprocessed emission is
conservatively set to the upper limit obtained previously ($f_{\rm{h}}=0.025$),
while the soft contribution (limited to $f_{\rm{s}}<0.03$) and the center and width
of the tophat were allowed to vary. This model provides a very poor fit to the data
($\chi^{2}_{\nu}=37/1$; $p\sim1\times10^{-9}$), severely underpredicting the
low frequency lag; the maximum lag produced by the model can be made similar
to that observed, but only by shifting the frequency at which this lag occurs
outside the range probed observationally. Allowing additional complexity in the
transfer function by allowing different tophat parameters for the soft and hard
bands does not resolve this discrepancy. Instead, a much larger hard band
contribution from the distant reflection is required to fit the lags, $f_{\rm{h}}>0.2$
(90\% confidence), roughly an order of magnitude greater than the upper limit
obtained from spectroscopy. Furthermore, even if the FeXXVI emission arises
from highly-ionised Compton reflection, such emission would contribute more
flux in the soft band than the hard, while the opposite relative contribution is
required to produce hard lags. We therefore conclude that any contribution from
distant reflection cannot explain the observed lags.

\section{Discussion and Conclusions}
\label{sec_dis}

We have presented a combined spectral and temporal analysis of the recent
\suzaku\ observations of the variable, unobscured AGN \ngc. Remarkably, the
X-ray spectrum of \ngc\ is well described simply with a basic powerlaw
continuum. The source does not display any evidence for the soft excess typically
displayed by other unobscured AGN (\citealt{Gierlinski04, Crummy06, Miniutti09,
Walton13spin}).

During this observation, \ngc\ displayed significant variability. In order to provide
a comparison with AGN that display more complex X-ray spectra, we investigated
the presence and frequency dependence of any lags between the emission
observed in the 0.5--2.0 and 2.0--5.0\,\kev\ energy bands, comparable to
recent studies undertaken for AGN that display clear soft excesses. Similar to
these sources, at low frequencies we find that the harder emission lags
significantly behind the softer emission, despite their substantially different soft
X-ray spectra (\citealt{McHardy04, FabZog09, Fabian13iras}). The natural
conclusion is that these hard lags are associated with the intrinsic powerlaw-like
continuum.

A variety of physical scenarios have been proposed to explain this energy- and
frequency-dependent temporal behavior when seen in other accreting black
holes: propagating accretion rate fluctuations (\citealt{Arevalo06}),
Comptonisation (\citealt{Kazanas97}) and distant reverberation (\citealt{Legg12}).
Of these, only the former two associate the hard lags with the intrinsic continuum,
while the latter associates them with a second emission component that arises
through reprocessing of the intrinsic continuum by relatively distant material.
However, we find that simple reverberation scenarios can be excluded, as the
contribution of any distant reprocessing in \ngc\ is too weak to produce the
observed lags. Furthermore, we stress that the same hard lags are observed in
Galactic BHBs, with sufficient data quality to allow detailed studies. The natural 
expectation is that these lags are produced by a common process. However,
\cite{Kotov01} also find the hard lags observed from Cygnus X-1 to be
inconsistent with a distant reverberation origin, due to the lack of the expected
iron feature in the lag energy-spectrum. Instead, \cite{Uttley11} argue strongly
in favour of the propagating fluctuations origin, based on the magnitude and
energy dependence of the lowest frequency lags observed from Galactic BHBs. We
therefore conclude this is also the likely origin of the lags in \ngc.

Although this dataset is not sensitive to high frequency variability, where the soft
lags are seen for sources \textit{with} soft excesses, our results most likely
support the hypothesis that the soft and hard lags arise through separate physical
mechanisms, as is strongly suggested by their differing lag energy-spectra
(\citealt{Zoghbi11, Kara13iras}). Indeed, our results are fully consistent with the
interpretation of the soft lags associated with the soft excess as the signature of
reverberation from the inner accretion disc, also implied by the more recent
detections of similar reverberation lags from the broad component of the iron line
in numerous AGN (\eg \citealt{Zoghbi13a}). In the few cases in which both the
iron line and the soft excess are observed (\ie bright, unobscured AGN), the lags
inferred independently from the soft excess and the broad iron line are fully
consistent (\eg \citealt{Kara13iras}), highly indicative that both features originate
from the same region of the accretion flow, and from a common physical process,
which is most likely relativistic disk reflection.

\section*{ACKNOWLEDGEMENTS}

The authors thank the reviewer for useful feedback. This research has made use
of data obtained from the \suzaku\ observatory, a collaborative mission between
the space agencies of Japan (JAXA) and the USA (NASA). CSR thanks support from
NASA under grant NNX12AE13G

\bibliographystyle{/Users/dwalton/papers/mnras.bst}

\bibliography{/Users/dwalton/papers/references.bib}

\label{lastpage}

\end{document}